\documentclass[mathleft %
]{an}
\usepackage{graphicx}
\usepackage{times}
\usepackage{color}
\usepackage{amsmath}

\newcommand{\prm}{Pm}
\newcommand{\rey}{Re}
\newcommand{\reym}{Rm}

\newcommand{\curl}{\vec{\nabla}\times}

\makeatletter
  \def\mathch{\protect\p@mathch}
  \def\p@mathch#1#2{%
    \begingroup
    \let\@nomath\@gobble \mathversion{#1}%
    \math@atom{#2}{%
    \mathchoice%
      {\hbox{$\m@th\displaystyle#2$}}%
      {\hbox{$\m@th\textstyle#2$}}%
      {\hbox{$\m@th\scriptstyle#2$}}%
      {\hbox{$\m@th\scriptscriptstyle#2$}}}%
  \endgroup}
  \def\bmath{\protect\p@boldm}
  \def\p@boldm#1{\mathch{bold}{#1}}
\makeatother

\newcommand\bE{\vec{E}}
\newcommand\bF{\vec{F}}

\newcommand\bb{\vec{b}}
\newcommand\be{\vec{e}}

\newcommand\bu{\vec{u}}

\newcommand\bnabla{\vec{\nabla}}

\newcommand\p{\partial}
\newcommand{\difft}[1]{\partial_t #1}
\newcommand{\f}[2]{\frac{#1}{#2}}

\newcommand\rmp{\mathrm{p}}

\newcommand{\rom}{R_\Omega}

\begin{document}

\Pagespan{1}{1}
\Yearpublication{xxxx}%
\Yearsubmission{xxxx}%
\Month{xxx}%
\Volume{xxx}%
\Issue{xxx}%

\title{Subcritical dynamos in shear flows}

\author{F. Rincon\inst{1,2}\fnmsep\thanks{Corresponding author:
  \email{rincon@ast.obs-mip.fr}\newline}
\and  G.I. Ogilvie\inst{2}
\and  M.R.E. Proctor\inst{2}
\and C. Cossu\inst{3}
}

\titlerunning{Subcritical dynamos in shear flows}
\authorrunning{F. Rincon, G.I. Ogilvie, M.R.E. Proctor, and C. Cossu}
\institute{
Laboratoire d'Astrophysique de Toulouse-Tarbes,
Universit\'e de Toulouse, CNRS, 14 avenue Edouard Belin,
F-31400 Toulouse, France
\and 
Department of Applied Mathematics and Theoretical Physics,
University of Cambridge, Centre for Mathematical Sciences, Wilberforce
Road, Cambridge CB3 0WA, United Kingdom
\and
Laboratoire d'hydrodynamique (LadHyX), CNRS -- \'Ecole
   polytechnique, 91128 Palaiseau cedex, France
}

\received{xxx}
\accepted{xxx}
\publonline{later}


\abstract{Identifying generic physical mechanisms responsible for the 
  generation of magnetic fields and turbulence in differentially rotating
  flows is fundamental to understand the dynamics of astrophysical objects 
  such as accretion disks and stars. In this paper, we discuss the concept
  of subcritical dynamo action and its hydrodynamic analogue
  exemplified by the process of nonlinear transition to turbulence in 
  non-rotating wall-bounded shear flows. To illustrate this idea, we
  describe some recent results on nonlinear hydrodynamic 
  transition to turbulence and nonlinear dynamo action in rotating shear flows 
  pertaining to the problem of turbulent angular momentum transport in accretion 
  disks. We argue that this concept is very generic and should be applicable to many
  astrophysical problems involving a shear flow and non-axisymmetric instabilities 
  of shear-induced axisymmetric toroidal velocity or magnetic fields,
  such as Kelvin-Helmholtz, magnetorotational, Tayler or global
  magnetoshear instabilities.  In the light of
  several recent numerical results, we finally suggest
  that, similarly to a standard linear instability, subcritical MHD
  dynamo processes in high-Reynolds number shear flows could act as
  a large-scale driving mechanism of turbulent flows that would in turn
  generate an independent small-scale dynamo.}

\maketitle

\section{Introduction}

\subsection{Differential rotation, non-normality and subcriticality}
Differential rotation is a major ingredient of the physics of
astrophysical objects and plays a central role in dynamo theory.
From a local point of view (zooming in on a narrow region 
of the flow), it can be decomposed into a global rotation
(Coriolis acceleration) and a shear flow (velocity gradient). Both
of these components are individually important for dynamo theory.
Let us focus on the shear component of differential rotation and
neglect global rotation for a moment. An important consequence
of the presence of shear is that it can induce a growth of velocity
field fluctuations (referred to in the hydrodynamic transition
community as the lift-up effect, see Sect.~\ref{sechydro}) or
magnetic field fluctuations (referred to in dynamo theory as 
the $\Omega$ effect for magnetic fluctuations having no spatial 
dependence along the direction of the shear velocity). 
These amplification processes, even though they result from linear 
terms in the equations, do not lead to an exponential growth of 
fluctuations on long timescales. Instead, velocity/magnetic
perturbations grow algebraically for a viscous/resistive timescale 
and  then decay if no other mechanism is present
to sustain them. The short-time dynamics associated with linear shear
flow operators is related mathematically to their non-normality
(Trefethen 1993; Schmid \& Henningson 2000), which makes it possible to combine
several individually decaying linear eigenvectors into a transiently 
growing structure that ultimately has to decay viscously or
resistively. The fact that transient growth is possible in shear flows,
combined with the observation that linearly stable shear flows become
nevertheless turbulent at moderate values of the Reynolds number, 
has given rise to the concept of subcritical, or bypass transition in
hydrodynamics. The word subcriticality here relates to the fact that 
transition in a linearly stable shear flow is possible at finite
values of the Reynolds number, while the critical Reynolds number for
the ``linear bifurcation'' is infinity in such a flow. In this subcritical
scenario, fluctuations that are transiently
amplified to finite amp\-li\-tu\-des can become linearly unstable, leading 
to sustained turbulence (or some form of complex nonlinearity in
general, see Baggett, Driscoll  \& Trefethen 1995) as a consequence of
the nonlinear saturation of the instability modes. 
Whether bypass transition  is possible or not is not completely
obvious, however, because one  has to ensure that the nonlinearities
that must come into play to  obtain sustained activity on long
timescales can actually play such a role (Waleffe 1995a). It was shown
by Hamilton, Kim \& Waleffe (1995) that  subcritical transition is
possible for non-rotating wall-bounded shear flows. 

In this paper, we aim at making use of the current knowledge on 
subcriticality in shear flows to understand some aspects of dynamo
action in differentially rotating astrophysical flows, thus we will
also consider some effects due to global rotation.
In the framework of this particular study, subcriticality will refer
to the observation that the sustenance of a given field relies on 
the presence of nonlinearity in the system, i.e. that if the system
was to be linearized around its background \textit{steady} state,
there would be no linearly unstable mode that could generate
turbulence or magnetic fields permanently. This definition has the
advantage of making it clear that subcriticality is related to the
dynamo problem.

\subsection{Subcriticality in accretion disks}
In the first part of the paper (Sect.~\ref{sechydro}-\ref{secmhd}), 
we will mostly concentrate on two problems pertaining to the issue of 
turbulent transport in accretion disks to illustrate subcriticality in 
rotating shear flows, which will enable us to make connexions with
different  types of dynamo problems involving differential rotation 
(Sect.~\ref{connexions}).

The first of these two problems is to understand the origin of turbulence 
in non-magnetized disks. The\-re is currently no known hydrodynamic instability, 
either linear or  non\-linear, in the  Keplerian shear flow regime
representative of the orbital dynamics of thin accretion  disks, that
could maintain the vigorous turbulent state required for accretion to
take place. The second problem is to understand the statistical
properties of turbulence in magnetized disks where a natural candidate 
for the generation of turbulence is the magnetorotational instability  (MRI, 
Velikhov 1959; Chandrasekhar 1960; Balbus \& Hawley 1991). The existence 
of a developed  MRI turbulence state relies on the presence of a sustained magnetic
field within the disk. If the central accreting object cannot provide an external
field threading the whole disk or if the disk resistivity is large enough 
(or the disk is not sufficiently ionized)  for fossil
fields to decay on short timescales in comparison to the disk
lifetime, the difficult question of the origin of turbulence in disks
directly translates into the equally difficult question of the
generation of magnetic fields - dynamo action - in these objects, and
its links with the MRI.

Let us first introduce the problem of turbulent transport in
\textit{non-magnetized disks}  or at least in cold and shielded
regions of disks  where the fluid is not coupled to magnetic 
fields, as in protoplanetary disks (e.g. Fromang, Terquem \& Balbus 2002).
The Rayleigh criterion tells us that a simple Keplerian shear
flow is linearly stable to axisymmetric perturbations from the
hydrodynamical point  of view, i.e. that turbulence cannot be
generated by a linear axisymmetric hydrodynamic instability 
in this flow. No local, linear non-axisymmetric hydrodynamic
instability is known either. Turbulence in a non-magnetized disk 
can therefore only result from a \textit{fundamentally
  nonlinear hydrodynamic process}. The existence of a subcritical 
transition to turbulence in linearly stable rotating shear flows
(particularly antiyclonic ones) has been invoked for many years
to explain the origin of turbulence in this context.
The main argument in favour of such a transition 
finds its roots in the well-known experimental evidence for transition to 
turbulence in  non-rotating wall-bounded shear flows (such as pipe
flow) that are also known to be linearly stable. This  argument is further
qualitatively strengthened by the observed high sensitivity of
shear flow stability to initial conditions, which has to do with the
previously mentioned non-normality of shear flow operators. 
There is a long ongoing debate on whether or not subcritical nonlinear 
transition is possible in Rayleigh-stable shear flows. A flavour of the
experimental debate can be found in Tillmark \& Alfredsson (1996), 
Richard \& Zahn (1999) and Ji et al. (2006), while on the numerical
and theoretical sides, we refer the reader to Hawley, Balbus \&
Winters (1999), Longaretti (2002) and Lesur \& Longaretti (2005).
A more exhaustive recent review  of this problem can be found in Rincon, 
Ogilvie \& Cossu (2007a). One of the purposes of the present paper is
simply to point out the important differences between the physics of 
subcritical transition in rotating and non-rotating shear flows 
in terms of transient growth and nonlinear interactions.

The problem of \textit{magnetized disks} is a priori a complete\-ly different 
one because of the existence of the MRI. In the presence of a mean field
threading the disk, the MRI grows velocity and magnetic field perturbations 
with optimally correlated radial and azimuthal components, providing a 
natural mechanism for angular momentum transport even in three-dimensional 
saturated regimes (but note that the efficiency of the process seems
to depend significantly on the magnetic Prandtl number (Lesur \&
Longaretti 2007)). The problem here is that there are large
uncertainties regarding the physical conditions that pertain to
different types and even different regions of disks.  It is in
particular not obvious that all disks are threaded by a mean field 
originating from the central
accreting object or that a preexisting fossil field can stay in the disk for 
a period of time comparable to a disk lifetime and sustain the MRI on
this timescale. Is it possible to obtain a sustained turbulent magnetized state 
in that case?  The recent discovery by Donati et al. (2005) that protoplanetary 
disks can host their own magnetic field, at least in their inner regions,
indicates that the question of the origin of magnetic fields in disks
is a relevant one. As mentioned earlier, answering this question
notably requires to understand if (MHD) dynamo action is 
possible in a Keplerian shear flow, which is a rather involved problem in view
of the current knowledge in dynamo theory. Let us try to make this point 
more evident, and assume that there is some undetermined form of dynamo 
action in a disk, which makes magnetic field perturbations grow in time. 
Such fluctuations are very likely to trigger the MRI locally very quickly
(the instability develops in weak-field regimes), showing the
imbrication of the dynamo process and the MRI. Besides, the process by
which the MRI works 
is the magnetic braking  of orbiting fluid particles, a fundamentally dynamical 
process that is not present in kinematic dynamos. Therefore, if dynamo action
is present in the system, it must be a \textit{fundamentally
  nonlinear  subcritical dynamo}. To cite Hawley \&
Balbus (1992), who first noticed that point, "the use of a kinematic 
dynamo model is inappropriate for an accretion disk [...] The turbulence is 
driven by the very forces [Lorentz forces] the kinematic dynamo excludes from 
the outset''. Even though this kind of dynamo looks fairly unusual,
Brandenburg et~al. (1995), Stone et~al. (1996) and Hawley, Gammie \& Balbus (1996)
demonstrated its existence. In particular, Hawley et~al. (1996)
compare the results of experiments in two and three dimensions, with
or without a Lorentz force term, to illustrate very clearly the
previous points. Some important 
limitations of these simulations have however recently been pointed out
(Fromang \& Papaloizou 2007a). If they do not challenge the existence
of the MRI dynamo process, at least at large magnetic Prandtl numbers
(Fromang et al. 2007b), they do however question its efficiency and
demonstrate  that how the MRI dynamo works remains a largely open
question that deserves further investigation.

\subsection{Outline}
The first goal of the present paper is to provide
connexions between the two forementioned hydrodynamic and MHD
problems, which are individually discussed in detail in 
Rincon et~al. (2007a) and  Rincon, Ogilvie \& Proctor  (2007b). 
As will be shown in Sect.~\ref{sechydro} and Sect.~\ref{secmhd}, 
this comparative approach
is motivated by several important similarities between the two
problems, namely hardwired nonlinearity and linear non-norma\-li\-ty.
In Sect.~\ref{sechydro}, we first discuss the dynamical principles
of transition in linearly stable shear flows and show that this problem
can be viewed as a nonlinear ``hydrodynamic dynamo'' problem
analogous to an MHD dynamo problem. This allows us to
point out the limits of the analogies between nonlinear hydrodynamic
instability in anticyclonic Rayleigh-stable shear flows and non-rotating
wall-bounded shear flows and to question the existence of a
subcritical transition in Keplerian flows. In Sect.~\ref{secmhd},
we make use of similar methods to address the MRI dynamo problem and
to illustrate the concept of the nonlinear self-sustaining MHD process.
We then move to the second major goal of this paper, which is to show
that the subcritical dynamo concept is a very
generic one that applies to many types of dynamo problems involving a
shear flow. This is done in Sect.~\ref{connexions}, where we also
discuss the relations between the subcritical dynamo model and more standard 
mean-field models such as the $\alpha\Omega$ dynamo. Finally,
based on these results and recent numerical results, notably
simulations of  MRI turbulence in the  dynamo regime, we suggest a 
possible scenario for the MRI dynamo  at large $\rey$ and $\reym$ 
involving small-scale dynamo action (Sect.~\ref{ssdynamo}). The
paper ends with a brief conclusion.

\section{Hydrodynamic transition in shear flows\label{sechydro}}
In this Section, we start by setting up a very simple hydrodynamic model problem, 
namely incompressible rotating plane Couette flow, to show that subcritical 
hydrodynamic transition to turbulence in some particular
differentially rotating flows can be analysed using dynamo
ar\-gu\-ments, i.e. in terms of dynamical exchanges between
``poloidal''  and ``to\-roidal''  components of velocity field
fluctuations with va\-nishing spatial average.
We then describe how transition to turbulence is actually thought to
work in non-rotating shear flows that are linearly stable for all
finite values of the Reynolds numbers using this dynamo
analogy\footnote{The term ``dynamo'' has already been
used in a hydrodynamic context to describe the generation of
large-scale vorticity fluctuations by small-scale helical turbulence 
(the AKA effect, see for instance Moiseev et~al. 1988), which is a distinct
phenomenon from what we describe in this paper.}. 
We finally build on  these results to discuss the possibility
of a similar hydrodynamic transition in linearly stable rotating shear
flows such as a Keplerian flow. The analysis performed here also sets
the scene for the MHD part of the paper and the following discussion.

\subsection{Incompressible rotating plane Couette flow}
\begin{figure}
\centerline{\includegraphics[width=80mm]{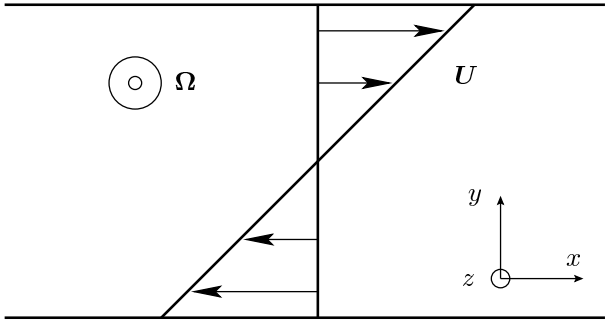}}
\caption{Geometry of rotating plane Couette flow.}
\label{figcouette}
\end{figure}

Incompressible rotating plane Couette flow, represented in Fig.~\ref{figcouette}, 
is a local approximation of differentially rotating flows such as those encountered 
in accretion disks. The flow is characterized by a background velocity field 
$\vec{U}=Sy\,\vec{e}_x$  with linear shear $S$ driven by countermoving walls 
situated at $y=\pm d$ and by a global, uniform in space, rotation vector 
$\vec{\Omega}=\Omega\,\vec{e}_z$. The Reynolds number for this flow is usually 
defined in the transition literature as 
\begin{equation}
  \label{eq:reynumber}
\rey=\frac{Sd^2}{\nu}
\end{equation}
 where $\nu$ is the kinematic viscosity. We can also define a rotation number,
\begin{equation}
  \label{eq:rotnumber}
  \rom=-\f{2\Omega}{S}~,
\end{equation}
which is positive for cyclonic flows (shear flow vorticity parallel to 
$\vec{\Omega}$) and negative for anticyclonic flows (shear flow vorticity anti-parallel 
to  $\vec{\Omega}$) and is related to a parameter commonly used in accretion disk theory 
to characterize differential rotation, $q=-d \ln \Omega/d\ln r$, 
accor\-ding to $\rom=-2/q$. A non-rotating flow has $\rom=0$, a flow
on the Rayleigh line has $\rom=-1$ and a Keplerian flow has
$\rom=-4/3$. The flow encounters an axisymmetric centrifugal instability 
for $-1<\rom<0$.
Our aim is to highlight the generic dynamo nature of this problem for
$\rom=0$ and $\rom=-1$ and its potential relevance to 
accretion disks and stars, so we choose to refer to the 
streamwise direction $x$ as being the toroidal direction and to the
$(y,z)$ plane as  being the poloidal plane. We further use the word
axisymmetric to qualify  $x$-independent 
perturbations (in the accretion disk terminology, $x$ corresponds to the azimuthal 
direction, $-y$ to the radial direction, and $z$ to the vertical direction).
Denoting by an overbar an average over $x$, any velocity field perturbation of the 
background flow is separated into an axisymmetric part $\bar\bu$ and a non-axisymmetric 
``wave'' part $\bu'$,

\begin{equation}
\bu=\bar\bu +\bu'\quad \mathrm{with}\quad \bnabla\cdot\bu=\bnabla\cdot\bar\bu=\bnabla\cdot\bu'=0.
\end{equation}

\noindent Note that both $\bar\bu$ and $\bu'$ are chosen to have zero
volume average, so that we have a form of dynamo problem for the
perturbations of the background Couette flow in the centrifugally
stable regimes. Using $d$ and $1/S$ as space and time units, the
momentum equation for the axisymmetric  toroidal $\bar u_x$ and 
poloidal $\bar\bu_\rmp$ components of the velocity field read
{\setlength{\mathindent}{0pt}
\begin{equation}
  (\difft{}+\bar\bu_\rmp\cdot\bnabla)\,\bar u_x+(\rom+1)\bar u_y=\f{1}{\rey}\Delta\bar u_x+
  F_x,
\label{uxbar}
\end{equation}}
{\setlength{\mathindent}{0pt}
\begin{equation}
  (\difft{}+\bar\bu_\rmp\cdot\bnabla)\,\bar\bu_\rmp-\rom\,\bar u_x\,\vec{e}_y=-\bnabla_\rmp\,\bar p\;+
  \f{1}{\rey}\Delta\bar\bu_\rmp+\bF_\rmp,
\label{upbar}
\end{equation}}
\noindent where

\begin{equation}
\label{eq:meanforce}
  \bF=-\overline{\bu'\cdot\bnabla\bu'}
\end{equation}

\noindent is the mean force associated with the non-axisymmetric part
of the flow and $p$ is the pressure divided by the constant density.
$\bar\bu_\rmp$ is a solenoidal two-dimensional velocity field that can
be written in terms of a streamfunction $\psi(y,z,t)$  with vanishing 
volume average:
\begin{equation}
\label{eqpsi}
  \bar\bu_\rmp=\bnabla\times(\psi\,\be_x)~.
\end{equation}
The associated toroidal vorticity reads
\begin{equation}
  \omega_x=-\Delta\psi.
\end{equation}
Taking the curl of Eq.~(\ref{upbar}) to eliminate
eliminate $\bar p$, one obtains

{\setlength{\mathindent}{0pt}
\begin{equation}
\label{eqtor}
  \difft{\bar u_x}-\f{\p(\psi,\bar u_x)}{\p(y,z)}= -(\rom+1)\bar u_y + \f{1}{\rey}\Delta\bar u_x+F_x~,
\end{equation}}
{\setlength{\mathindent}{0pt}
\begin{equation}
\label{eqpol}
  \difft{\omega_x}-\f{\p(\psi,\omega_x)}{\p(y,z)}=-\rom\p_z\bar u_x+\f{1}{\rey}\Delta\omega_x+\vec{e}_x\cdot{\curl{\bF}}~
\end{equation}}

\noindent where $\p(,)/\p(,)$ denotes the Jacobian. It can be seen that both 
$\bar u_x$ and $\omega_x$ have linear source or sink terms proportional
to $\bar u_y$ and $\p_z\bar u_x$, and a fully nonlinear 
source or sink term associated with the axisymmetric part of the
advection term $\vec{F}$. This observation also stands for shear
profiles different from that of  plane Couette flow.

\subsection{Subcritical transition in non-rotating shear flows}
Let us now specialize to interesting particular cases and first concentrate on the 
non-rotating case $\rom=0$. The phenomenology of transition in non-rotating shear 
flows has been a mystery since the end of the $19^\mathrm{th}$ century
and the experiments by O. Reynolds (1883). For a long time, the
experimental observation that wall-bounded
shear flows encounter transition to turbulence at modest $\rey$ values could not be 
reconciled with theory, which could only predict the absence of linear instability in 
many of these flows. There is now some significant evidence that the 
transition is related to a nonlinear mechanism christened the self-sustaining 
process (SSP), which was first described in detail by Hamilton et~al. (1995).
 The actual complexity of this transition and its sensitivity to initial 
conditions depends on the background shear profile (e.g. pla\-ne Couette flow, 
plane Poi\-seuil\-le flow, pipe Hagen-Poiseuille flow or Blasius boundary layer) and 
on the structure of the phase space of the corresponding
dynamical system (i.e. the presence of fixed points, homoclinic or
heteroclinic orbits, etc.). Tran\-si\-tion  is currently best
understood for plane Couette flow, which encouters a simple
saddle-node bifurcation (Rincon et~al. 2007a; Wang, Gibson \& Waleffe
2007; Viswanath 2007), while pi\-pe Ha\-gen-Poiseuille flow, for
instance, seems to encounter a complex chaotic transition
whose limits in parameter space are now referred to as
``the edge of chaos'' (Schneider, Eckhardt \& Yorke 2006, 2007).

A nice way of describing the SSP is to think of it in terms of
subcritical ``hydrodynamic  dynamo''. We first note that for $\rom=0$,
only $\bar u_x$ has a linear ``source''  term. This term is a
hydrodynamic analogue of the $\Omega$ effect of dynamo theory. It
is actually called the lift-up effect in the transition literature 
(Ellingsen \& Palm 1975; Landahl 1980) and is of course a purely
axisymmetric effect which leads to algebraic linear amplification of 
the toroidal velocity field. However, as for the $\Omega$ effect,
this is only a transient effect, for in the absence of a poloidal 
source term to regenerate the poloidal velocity field, viscous damping 
ultimately kills both poloidal and toroidal velocity fields on a 
viscous time scale. As can be seen in Eq.~(\ref{eqpol}) there is no 
linear poloidal velocity source term in the absence of rotation,
so that the only way poloidal motions can be sustained is via the 
nonlinear interaction term $\curl{\bF}$, which is only non-vanishing
if the total flow has a non-axisymmetric part. This is clearly a form
of antidynamo theorem for non-rotating linearly stable 
hydrodynamic shear flows. Therefore, the dynamo question is to ask how 
non-axisymmetry can emerge in  such a system, leading to a
self-sustaining solution via nonlinear feedback
on poloidal motions. To answer this question, one has to look at what 
the lift-up effect actually produces. Starting with a
$\mathcal{O}(1/\rey)$ poloidal velocity field depending on $y$ and
$z$, the lift-up effect has the ability to generate on a 
$\mathcal{O}(\rey)$ timescale a toroidal velocity field also depending on 
$y$ and $z$ with an $\mathcal{O}(1)$ amplitude, comparable to that of
the background shear 
flow. There is a lot of experimental evidence for this so-called
``streaks'' field (Hamilton et~al. 1995). As this field is modulated
in $z$, it is actually a finite-amplitude shear flow that
contains inflexion points in that direction, and is therefore unstable 
to non-axisymmetric Kelvin-Helm\-holtz type instabilities. 
Waleffe (1995b, 1997 1998) showed that the three-dimensional velocity field
associated with these instabilities leads to an $\bF$ term that can
regenerate the weak poloidal velocity field. In other words,
Reynolds stres\-ses  generated by the secondary instability in its
weakly nonlinear regime can close the dynamo loop. For a given $\rey$
value, it is possible to show that there is a continuous family of 
self-sustained solutions with different aspect ratio, i.e. different
periodicity in $x$ and $z$.

A very important requirement for the process to work is that there must
be a very good spatial overlap (correlation) between the nonlinear 
interaction term $\bF$ and the axisymmetric poloidal velocity. The spatial 
shape of the feedback is notably very sensitive to the geometry of the 
flow and to the wavelength of the instability mode. A second important 
remark is that the process is fully nonlinear and subcritical for two
reasons. First, for a given $\rey$, a finite amplitude
$\mathcal{O}(1/\rey)$ poloidal field is required to initiate the process.
Such an amplitude is of course extremely small at very large $\rey$
but can never be taken infinitesimally small as would be the case for
a linear instability  (Nagata (1990) described the corresponding
branch of nonlinear solutions - which had not  been described in terms
of a SSP at that time - as a ``bifurcation from infinity''). 
The second reason is that the feedback term is a fully nonlinear
interaction term due to the non-axisymmetric instability mode. To
summarize the process in terms of an initial value problem, let us 
enumerate the three elements that are required to obtain a 
subcritical ``hydrodynamic dynamo'' in a non-rotating shear flow:

\smallskip

\begin{enumerate}
\item linear transient amplification up to $\mathcal{O}(1)$ amplitudes of an 
axisymmetric toroidal velocity field, from a seed but finite 
 $\mathcal{O}(1/\rey)$ axisymmetric poloidal velocity field,
\item[]
\item non-axisymmetric linear shear instability of the finite-amplitude,
$z$-modulated axisymmetric toroidal velocity field,
\item[]
\item regeneration of the weak axisymmetric poloidal velocity field by 
nonlinear self-interactions of the non-axisym\-metric instability mode.
\end{enumerate}

\noindent An equivalent description of the SSP can be given in terms of a coherent 
structure (Rincon et~al. 2007a) which, for pla\-ne Couette flow, is a saddle
fixed point in phase space. The initial value problem formulation described a\-bo\-ve
then consists in following  phase space trajectories sticking as much as possible
to the stable manifold of the fixed point and approaching it closely before being 
ejected along its unstable manifold, towards a turbulent attractor. A very nice 
graphical representation of this process is given by Gibson, Halcrow and 
Cvitanovi\'c (2008).

\subsection{Subcritical transition in rotating shear flows}
Is a subcritical hydrodynamic transition possible in linearly stable
\textit{rotating} shear
flows? As mentioned in the introduction, the observation that their non-rotating
linearly stable counterparts become turbulent at modest values of 
$\rey$ has been used as an argument in favour of such a scenario for a long time.
Now that transition in non-rotating flows is better understood in terms of the SSP,
a natural question to ask is whether a similar process can occur in rotating shear flows.
This question was addressed by Rincon et~al. (2007a). Here, we summarize the main points 
of this study using the subcritical dynamo phenomenology. The starting point is that the 
linearized non-rotating case has an interesting linearly stable rotating counterpart, 
the Rayleigh-line regime $\rom=-1$ (we note in passing that the Rayleigh-line regime is 
similar to the Keplerian regime in the sense that both regimes are
anticyclonic and linearly  stable. The Rayleigh line is of course
closer to the centrifugal instability region). For $\rom=-1$, the linear source
term in Eq.~(\ref{eqtor}) is exactly vanishing, but there is still a
linear Coriolis acceleration term in Eq.~(\ref{eqpol}). From the linear 
point of view, this means that the lift-up effect has an exact analogue
on the Rayleigh line. This effect, christened the anti lift-up effect
by Antkowiak \& Brancher (2007), has the ability to transiently
generate a finite-amplitude axisymmetric poloidal velocity field from
a seed $\mathcal{O}(1/\rey)$ axisymmetric toroidal velocity field.
In other words, streaks now generate rolls.  Note that this effect can
be understood as an epicyclic oscillation with infinite period
(on the Rayleigh line, the epicyclic frequency is zero) leading to a
growth of toroidal vorticity proportional to time (Antkowiak 2005,
Antkowiak \& Brancher 2007). The fact that such an axisymmetric linear
transient amplification effect is present on the Rayleigh line 
is certainly encouraging for the prospect of a rotating SSP, but the full picture 
requires nonlinearity to come into play. In that respect, the Rayleigh
line regime appears to be far less favourable than the non-rotating
case, for two important reasons. The first one is that the
presence of a finite-amplitude axisymmetric poloidal flow on the
Rayleigh line, unlike that of the axisymmetric toroidal flow in the 
non-rotating case, causes a strong poloidal nonlinear advection of
the total axisymmetric flow via the nonlinear terms after $\p_t$ in
Eqs.~(\ref{uxbar})-(\ref{upbar}) as soon as the poloidal flow reaches a significant 
amplitude by means of the anti lift-up effect. This renders the axisymmetric part
of the flow much more complex than in the non-rotating case. The
second reason is that the  non-axisymmetric instabilities of the
resulting nonlinear axisymmetric flow do not generate a nonlinear
feedback that would show a good spatial correlation with the seed 
streaks field. These instabilities are actually quite different from
those encountered in the non-rotating case for now the axisymmetric
flow is dominated by its poloidal component instead of its toroidal
component which was unstable to shearing instabilities in the non-rotating
regime. Overall, we found it impossible to proceed as in the
non-rotating case to obtain 
fully nonlinear three-dimensional steady solutions of the fluid equations.

\begin{figure}
\centerline{\includegraphics[width=80mm,height=50mm]{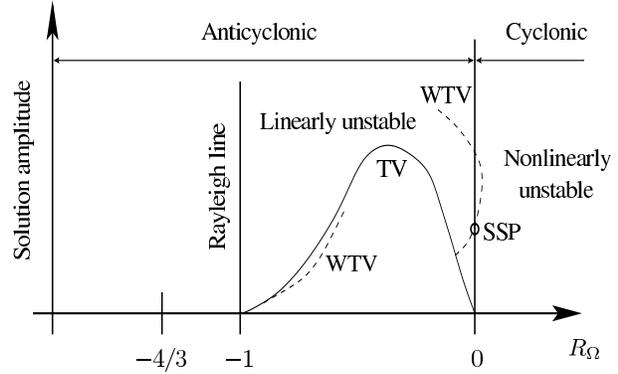}}
\caption{Stability diagram for rotating plane Couette flow for a value 
of $\rey$ in the domain of existence of the self-sustaining process.
The amplitude of the solutions in this diagram is somewhat arbitrary
but is intended to approximate the behaviour of the shearing rate 
at the flow boundaries. The TV full line curve is for nonlinear 
axisymmetric Taylor Vortices arising in the linearly unstable
centrifugal regime, the dashed WTV lines is for the three-dimensional
Wavy Taylor Vortices  bifurcating from the axisymmetric TVs, and SSP
stands for the fixed point  associated with the self-sustaining
process at $\rom=0$. This point can be
obtained by direct continuation of WTVs to $\rom=0$ and the
corresponding branch of solutions can be continued in the linearly
stable cyclonic region (see Rincon et~al. (2007a) for further details
and references).}
\label{figstab}
\end{figure}

The current status regarding stability with respect to $\rom$ in
rotating plane Couette flow is depicted in Fig.~\ref{figstab}. Several 
comments on this figure are in order. First, it is important to note
that subcritical solutions
obtained in the non-rotating case can be continued nonlinearly into
the anticyclonic linearly unstable regime, where they take on the form of 
nonlinear wavy Taylor vortices (instabilities of the Taylor vortices
that result from the centrifugal instability in this regime). They can
also be continued into the cyclonic linearly stable regime which is
therefore nonlinearly unstable. A similar behaviour cannot be found
in the neighbourhood of the Rayleigh line,
demonstrating that a symmetry between the cyclonic regime and its
anticyclonic counterpart beyond the Rayleigh line does not exist in
three dimensions (the absence of this symmetry was also demonstrated
by Lesur \& Longaretti (2005) using direct numerical
simulations).  The second comment is that in the case of a general
differential rotation  profile for which $\rom$ is different from $0$ or $-1$, 
there is no lift-up or anti lift-up effect anymore, and that
axisymmetric algebraic growth turns into epicyclic oscillations with
a non-zero frequency. In the Keplerian case, only non-axisymmetric
algebraic transient growth (also sometimes called swing amplification)
is possible,  which makes it pretty much useless to think in terms of
a po\-loidal and toroi\-dal  description of the hydrodynamic
problem. The only way the Keplerian regime 
could be related to the previous type of subcritical hydrodynamic
dynamo would be to make a nonlinear connexion with a SSP on the
Rayleigh line which, as we showed, probably does not exist (Rincon et
al. 2007a). It therefore appears that subcritical transition in
anticyclonic shear flows, if any,  has only very little to do with
subcritical transition in non-rotating
or cyclonic shear flows. The possibility of a subcritical dynamo
mechanism relying on instabilities of transiently amplified
non-axisymmetric structures remains open but recent high-resolution
simulations have not been able to isolate such a physical process
either (Shen, Stone \& Gardiner 2006).

\section{Subcritical MRI dynamo in Keplerian flow\label{secmhd}}
We now move to the MHD problem of uncovering the physical processes that give
rise to what is called MRI dynamo action in Keplerian shear flows. For
this purpose, we keep our rotating plane Couette flow description,
setting $\rom=-4/3$. This problem has much in common with the
non-rotating hydrodynamic problem, which makes it appealing to attempt
to apply the SSP phenomenology again. First, as mentioned in the introduction,
any MRI dynamo must be intrinsically nonlinear, because the Lorentz force is mandatory 
for the MRI. Besides, since magnetic fields feel shear but not
Coriolis acceleration, it is possible to  algebraically (and
transiently) amplify axisymmetric toroidal fields from poloidal 
fields through the $\Omega$ effect in the same manner as the amplification of 
streaks from rolls in the non-rotating hydrodynamic problem (see
Livermore \& Jackson (2004) for a discussion on non-normality in
MHD). One might then ask if the toroidal magnetic field  encounters
non-axisym\-metric instabilities such as an MRI and if the nonlinear
interactions of such instabilities have the ability to close the
dynamo loop by feeding back on the axisymmetric poloidal field. 

It is straightforward to show that the induction part of the MHD
rotating plane Couette flow problem takes on a form very similar to
the momentum equation for the non-rotating  hydrodynamic shear flow
problem. For this purpose, we introduce the magnetic Reynolds number
\begin{equation}
  \label{eq:reym}
  \reym=\f{Sd^2}{\eta}~,
\end{equation}
where $\eta$ is the magnetic diffusivity, and we decompose the
magnetic field $\bb$ in terms  of an axisymmetric part $\bar\bb$ and a
non-axisymmetric wave part $\bb'$
\begin{equation}
\bb=\bar\bb +\bb'\quad \mathrm{with}\quad \bnabla\cdot\bb=\bnabla\cdot\bar\bb=\bnabla\cdot\bb'=0~.
\end{equation}
Note that both $\bar\bb$ and $\bb'$ are chosen to have zero volume
average, so that we are dealing with a genuine dynamo problem (no net
magnetic flux is threading the domain).  Similarly to the hydrodynamic
problem, we introduce a flux function $\chi(y,z,t)$ to  describe the
axisymmetric part of the poloidal magnetic field $\bar\bb_\rmp$
\begin{equation}
  \label{eq:bpol}
\bar\bb_\rmp=\curl{\left(\chi\vec{e}_x\right)}~.
\end{equation}
The induction equation for $\chi$ and for the axisymmetric to\-roidal
component of the magnetic field $\bar b_x$ reads

{\setlength{\mathindent}{0pt}
\begin{equation}
\label{eq:induc1}  
 \difft{\bar  b_x}-\,\vec{e}_x\cdot\curl{(\bar\bu\times\bar\bb)}=\bar b_y+\,
\vec{e}_x\cdot\curl{\bE}+\f{1}{\reym}\Delta\,\overline{b}_x~,
\end{equation}}
{\setlength{\mathindent}{0pt}
\begin{equation}
 \label{eq:induc2}  
 \difft{\chi}-\f{\partial\,(\psi,\chi)}{\partial\,
 (y,z)}=E_x+\f{1}{\reym}\Delta\,\chi~,
 \end{equation}}
\noindent\!\!\! where $\psi$ has been defined in Eq.~(\ref{eqpsi}), the
 first linear term on the right hand side of Eq.~(\ref{eq:induc1})
 stands for magnetic induction by the shear flow and
\begin{equation}
  \label{eq:emf}
  \bE=\overline{\vec{u'}\times\vec{b'}}
\end{equation}
is the axisymmetric part of the electromotive force (EMF) generated by
the nonlinear interaction between the wave parts of the velocity and
magnetic fields. It is straightforward to notice that
Eqs.~(\ref{eq:induc1})-(\ref{eq:induc2}) share the most interesting
characteristics  of Eqs.~(\ref{eqtor})-(\ref{eqpol}), namely linear
non-normali\-ty in the toroidal part and nonlinear  feedback of
three-di\-men\-sion\-al structures in the poloidal part. Taking once
again the point of view of an initial value formulation and starting
from an  axisymmmetric $\mathcal{O}(1/\reym)$ poloidal magnetic field
(expressed in terms of an Alfv\'en velocity), an axisymmetric $\mathcal{O}(1)$
toroidal field can be transiently generated on a timescale
$\mathcal{O}(\reym)$ by the $\Omega$ effect.
Rincon et~al. (2007b) showed (solving the  full MHD equations,
including a Lo\-rentz force  in the momentum equation) that such a
toroidal field encounters a non-axisymmetric instability. There are
several instability modes with different symmetry properties which
can be predicted from the symmetries imposed on the axisymmetric
fields. The important point is that for some of  these modes,
nonlinear  interactions give rise to a toroidal EMF which 
axisymmetric  projection has the ability to regenerate the initial
poloidal seed field, thereby leading to  dynamo action. We interpreted
these instabilities in terms of an MRI of the axisym\-me\-tric
toroidal field: the axisymmetric poloidal field here is
too weak to be MRI-unstable (only very large poloidal wavenumbers
would be unstable, and they are damped by viscous and resistive terms
in our dissipative set-up). Another indication is that the
instabilities are purely non-axi\-sym\-metric, which is in line with
the MRI instability growth rate $\gamma$ 
obtained from the local dispersion relation $\gamma\sim\vec{k}\cdot\vec{V}_A$
where the Alfv\'en velocity $\vec{V}_A$ would be dominated by the
toroidal field. Also, the modes are spatially centred and symmetric
or antisymmetric with respect to  the local extrema of the toroidal
field in the poloidal plane. Finally, we noticed that the presence of
the instability required an MRI unstable rotation to be present.

Overall, we found it possible to obtain nonlinear fixed points in the
same way as in the hydrodynamic problem, but only for a very
restricted range of low values of $\rey$. In this regime,
non-axisymmetric modes have  real eigenvalues corresponding to purely
imaginary frequencies (the modes arise from a steady pitchfork
bifurcation).  The full three-dimensional nonlinear solution that can
be computed from these modes therefore takes the form of a steady 
solution (equivalently a fixed point). For larger values of $\rey$,
we observed collisions between pairs of real eigenvalues of the
instability modes, which turn into complex conjuguate pairs (the
corresponding modes subsequently arise from a Hopf bifurcation). In
such a situation, we expect nonlinear travelling waves to be 
present instead of steady solutions, similarly to what happens in the
hydrodynamic problem for Hagen-Poiseuille flow (Wedin \& Kerswell
2004) or plane Poiseuille flow (Waleffe 2001). These travelling waves
can in principle be captured numerically exactly like steady solutions
(by performing a Galilean transformation and adding the phase speed as
an extra unknown in the problem). In the present MHD problem, however,
such a numerical solution cannot be obtained easily because a
symmetry used to decrease the computing costs is lost when the mode
eigenvalues turn into complex conjugate pairs. Seeking coherent 
time-dependent solutions to the subcritical MRI dynamo problem using
direct time-stepping techniques therefore looks more promising than
using continuation techniques such as those described in Rincon
et~al. (2007a, 2007b). Such an investigation is currently 
underway (Lesur \& Ogilvie 2008).

Like the hydrodynamic self-sustaining process, the process described
above is genuinely nonlinear. First, a small but finite-amplitude
seed axisymmetric poloi\-dal
field is required to obtain a suffi\-cien\-tly large axisymmetric 
toroidal field to trigger non-axisymmetric instabilities (but
sufficien\-tly weak at the  same time for an MRI to be
possible). Besides, two nonlinear terms, a Lorentz force and a 
fluctuating EMF, are needed for the instability to develop and for the
feedback on the axisymmetric poloidal  field to be possible. Note that
the Lorentz force is a nonlinear term in that context because the
axisymmetric toroidal field that becomes MRI-unstable is part of the
total magnetic field perturbation (it has zero net flux and would
decay on a resistive timescale  without the three-dimensional
instability feedback). To summarize this self-sustaining MHD process
by means of an initial value description, let us restate the three
elements that are required to obtain a subcritical MRI dynamo in a
Keplerian shear flow:

\smallskip

\begin{enumerate}
\item linear transient amplification of an axisymmetric toroi\-dal magnetic field by 
the $\Omega$ effect acting on a $\mathcal{O}(1/\reym)$ seed axisymmetric poloidal 
magnetic field,
\item[]
\item non-axisymmetric linear instability (MRI) of the finite-amplitude axisymmetric 
toroidal magnetic field,
\item[]
\item regeneration of the seed axisymmetric poloidal mag\-ne\-tic field by 
nonlinear self-interactions of the non-axi\-sym\-me\-tric instability mode.
\end{enumerate}

\noindent As discussed just before, an equivalent description of this SSP can be done 
in terms of steady or travelling coherent structures (Rincon et~al. 2007b). A tentative 
interpretation of the role of such coherent structures in simulations of MRI turbulence 
in zero-net-flux set-ups is suggested in Sect.~\ref{ssdynamo}.

\section{Connexions with other dynamo models\label{connexions}} 
The phenomenology of the subcritical shear dynamo scenario presented
abo\-ve is obviously not specific to Keplerian shear flows. 
Dynamos models relying on non-axisymmetric  hydromagnetic
instabilities in differentially rotating flows 
have been thought for in the context of the geodynamo 
for instance (Fearn \& Proctor 1983, 1984, 1987). In this Section, we show
that several recent dynamo models introdu\-ced in astrophysics by Cline, 
Brummell \& Cattaneo (2003), Spruit (2002) and Miesch (2007a)
fit perfectly into the subcritical dynamo concept. We then attempt to
compare the subcritical scenario with more standard scenarios such
as the $\alpha\Omega$ dynamo and to provide some connexions between
these two views. Finally, we comment briefly on another form of
subcritical dynamo recently discovered by Ponty et~al. (2007).

\subsection{Dynamos driven by non-axisymmetric instabilities}
A fundamental ingredient in the subcritical dynamo loop pre\-sen\-ted in the previous
sections is the possibility of triggering non-axisymmetric instabilities in the system. 
Working in the Keplerian regime  allows for one such instability, the MRI, to 
exist, while the subcritical hydrodynamic transition in non-rotating
shear flows relies on a non-axisym\-me\-tric  instability triggered by
a spatially modulated shear flow profile.  The interesting point is
that there are many other sheared systems where non-axisymmetric MHD or
hydrodynamic instabilities of shear-induced axisymmetric toroidal fields 
can develop. Here, we will focus on several
dynamo models that have recently been discussed in the astrophysical
literature (we will come back to the previously mentioned geodynamo model
of Fearn \& Proctor (1984, 1987) in the next paragraph). An
interesting example is that of the shear-buoyancy dynamo discovered
by  Cline et~al. (2003) using direct numerical simulations (a typical
initial value problem), which can be summarized in the following way:

\smallskip

\begin{enumerate}
\item linear transient amplification of an axisymmetric toroi\-dal magnetic field by 
the $\Omega$ effect acting on a seed axisymmetric poloidal magnetic field,
\item[]
\item axisymmetric buoyancy instability of the toroidal field inducing an axisymmetric 
poloidal flow,
\item[]
\item linear transient amplification of an axisymmetric toroi\-dal velocity field from
the axisymmetric poloidal flow (thereby steepening the original shear),
\item[]
\item non-axisymmetric linear instabili\-ty (Kelvin-Helmholtz) of the steeper
axisymmetric toroidal velocity field,
\item[]
\item regeneration of the seed axisymmetric poloidal mag\-ne\-tic field by 
nonlinear self-interactions of the non-axi\-sym\-me\-tric instability mode.
\end{enumerate}

\noindent This loop looks very similar to the subcritical dynamo loops described 
earlier and indeed Cline et~al. (2003) clearly noticed that their
dynamo was intrinsically nonlinear. Several other elements in their
paper further demonstrate that these dynamos proceed in the
same way. First, Cline et~al. (2003) mention a poloidal magnetic field
threshold depending on $\reym$ under which no dynamo action is
possible, very much like in the Keplerian problem (in which the
threshold amplitude scales like $1/\reym$) and in the non-rotating hydrodynamic 
problem (a $1/\rey$ threshold amplitude is possible in that case
but this point is still a matter of debate though, see Kreiss, Lundbladh \&
Henningson (1993), Baggett \& Trefethen (1997), Chapman (2002) and the
experiments by Peixinho \& Mullin (2007)). Since the threshold is determined by the
physics of algebraic  growth via the $\Omega$ effect, which is present in all these
problems, we suspect that an axisymmetric poloidal magnetic field amplitude
scaling like $1/\reym$ is also required in the shear-buoyancy case 
for the dynamo to be triggered. Cline et~al. (2003)  also  report
stea\-dy, cyclic and chaotic regimes depending on their parameters, which is in
line with our previous argument regarding the possibility of having
time-dependent solutions in the large $\rey$ regime of the Keplerian
problem and the discovery of an ``edge of chaos'' in hydrodynamical
systems. We finally note that an important  reason why the Cline
et~al. (2003) dynamo works is because they deal with a non-rota\-ting
system, meaning that the poloidal velocity field
perturbations associated with the magnetic buoyancy instability are
getting wound up into strong toroidal velocity field perturbations by
the lift-up effect (step 3 of their process). It is that  toroidal
velocity field that renders the total shear profile unstable to
non-axisym\-metric Kelvin-Helm\-holtz 
instabilities, which in turn produce the feedback. As mentioned in
Sect.~\ref{sechydro}, it is not possible anymore to produce a strong
axisymmetric toroidal velocity field  and the associated
Kelvin-Helmholtz instability in the presence of a global rotation rate 
comparable to the shearing rate. Their dynamo might be continued into
a weakly rotating regime like the hydrodynamic SSP, but its existence in 
regimes with comparable shearing and rotation timescales is more
difficult to predict.

Another problem which takes on the same form is that of dynamo action in stellar 
radiation zones. Spruit (2002) suggested  the existence of a dynamo  relying on the 
non-axisymmetric Tayler (1973) - Pitts \& Tayler (1985) instability of toroidal 
magnetic fields generated by the $\Omega$ effect in that context. 
This scenario seemed to be comforted  by numerical simulations
performed by Braithwaite (2006), but this view has recently 
been challenged by Zahn, Brun \& Mathis (2007) and {Gellert},
{R\"udiger} \& Elstner (2008)
on the basis of direct numerical  simulations. Zahn et~al. (2007) do observe the 
amplification of an  axisymmetric toroidal field  from a seed poloidal field via 
the $\Omega$ effect and the subsequent occurrence of a non-axisymmetric instability, 
exactly like in the previous dynamo problems, but they fail to obtain a  significant 
feedback of the instability on the axisymmetric poloidal field, whose
fate is therefore resistive 
decay. The critical point in the scenario is therefore how the
feedback is produced. Spruit (2002)
and Braithwaite (2006) argued that the feedback is done directly by an
energy exchange  between the non-axisymmetric instability mode and
either the axisymmetric poloidal or toroidal field. This argument has
been criticized by Zahn et al. (2007), 
who pointed out that the only way to obtain a feedback from an $m=1$ component into an 
$m=0$ component was through nonlinear interactions. So, the current
problem to understand whether this dynamo can operate is to determine
whether such a regenerating nonlinear feedback is 
possible in the system. Our experience with the MRI problem and the
hydrodynamic problem is that this is extremely sensitive to the
spatial correlation between the nonlinear feedback term and
the original axisymmetric poloidal field. This correlation is quite
sensitive to the aspect ratio of the 
instability mode, which depends on the poloidal localization of the mode and 
its non-axisymmetric wavenumber. In the simulations by Zahn et~al. (2007), the 
instability is of course localized  where the toroidal field is strong, i.e. in a 
very narrow zone of large differential rotation,
which covers a small poloidal area compared to the total poloidal area
covered by the  poloidal magnetic field introduced originally. A more
favourable situation for the
Spruit dynamo to be found (but not necessarily a more realistic one
from the astrophysical  point of view!) probably requires a large
poloidal overlap between the differential rotation  zone and the
axisymmetric poloidal field. This way, the resulting non-axisymmetric
instability of the toroidal field generated through the $\Omega$
effect would probably cover a larger poloidal area and feedback
more coherently on the axisymmetric poloidal magnetic field.

We finally briefly mention that another astrophysical situation in
which a subcritical shear dynamo could be operating is that of the
solar tachocline. In the model of Miesch (2007a), the imposition of
a latitudinal shear and the occurrence of non-axi\-sym\-metric global
magnetoshear instabilities (Miesch, Gil\-man \& Dikpati 2007b) like
the clamshell instability provide all the necessary ingredients for
such a dynamo.

\subsection{$\boldsymbol{\alpha\Omega}$ and kinematic dynamos}
There are clearly major differences between the subcritical dynamo scenario and 
a mean-field $\alpha\Omega$ dynamo scenario. The first one is the absence of
kinematic regime in the subcritical case (hence its name). The second
one, which is more qualitative, is that subcritical dynamos discovered
so far take on the form of three-dimensio\-nal struc\-tures that are
extremely coherent in both time and space, as shown spectacularly by
Cline et al. (2003),  while mean-field theory relies on a statistical
description of the dynamo process. Specific differences between the
$\alpha\Omega$  dynamo and the MRI dynamo have also been pointed out
by Hawley \& Balbus (1992) and Brandenburg et~al. (1995). 

So, is there a way to reconcile both views somehow? One can of course
attempt to envision the nonlinear feedback of non-axisymmetric
instability modes as some form of mean-field back reaction on the
poloidal magnetic field (see for instance Zahn et al. (2007) and
Gellert et al. (2008) for the case of the Tayler instability).
A possible route towards a unified description is given
by Moffatt (1970), who attempted to compute an $\alpha$ from an
ensemble of inertial waves. In his model, each individual wave gives
rise nonlinearly to a small average EMF. A random ensemble of such waves
can generate a net $\alpha$ effect provided that the velocity 
field associated with the waves lacks reflexional symmetry. Similarly,
one can attempt to calculate an $\alpha$ effect or at least a toroidal
EMF from a collection  of non-axisymmetric instabilities, or to solve
simultaneously for an axisymmetric mean-field model and a single-mode
non-axisymmetric instability, taking the EMF produced by the
non-axisymmetric instability as an input for the mean-field model. 
Such a so-called $2\frac{1}{2}$D model was introduced in the
context of the geodynamo by Fearn \& Proctor (1984, 1987).
In their case, the non-axisymmetric instability is a convective
instability in a differentially rotating sphere (Fearn \& Proctor
1983), so once again all the ingredients for a subcritical shear
dynamo are present in this problem. They failed to obtain dynamically
consistent steady dynamo solutions using an iterative solver but
Jones, Longbottom \& Hollerbach (1995), following the same
idea, found time-dependent solutions to the same problem using direct
time-stepping. 

We note finally that Cline et al. (2003) showed that the feedback
process in their dynamo loop could not be cast in the simple
mathematical form of an $\alpha$ term, so it is not clear currently
whether one can actually construct a standard mean-field model from a
subcritical dynamo in general. Also, depending on the symmetries of
the problem, it is possible that some modes exert some destructive
feedback instead of a regenerating one, so that the subcritical
dynamo effect could disappear on average in some cases. A way to avoid
this kind of cancellation on average is to impose some form of global
symmetry breaking in the system, like a global rotation of the
system. In contrast, a coherent dynamo process consisting of
individual events in time (like the buoyant rise of individual
magnetic flux tubes) does not require this kind of ingredient. For instance, 
the recurrent bursting events during which turbulence gets generated
all of a sudden in otherwise quiescent shear flows (some form of
time-dependent ``hydrodynamic dynamo'') have often  been associated
with the hydrodynamic SSP  (Waleffe 1997; Jimenez \& Pinelli 1999;
Jimenez \& Simens 2001) which in its simplest form does not rely on
any kind of imposed symmetry-breaking.

\subsection{Subcritical dynamo in the Taylor-Green flow}
Before we close this section, we briefly discuss another type of subcritical
dynamo discovered recently by Ponty et~al. (2007) using a three-dimensional 
Taylor-Green forcing for the velocity field.  A fundamental difference 
between their problem and the problem discussed here is that we do impose a 
global shear while they only have local velocity gradients in their flow. 
An important consequence of imposing a global shear in
the system is that the subcritical dynamo branches bifurcate ``from 
infinity'' (i. e. they asymptote to zero as the control parameter -
$\reym$ here - tends to infinity, see Nagata (1990)) while in their
problem, there is a real linear dynamo bifurcation with a well-defined
critical value for $\reym$. Their dynamo
is subcritical in the sense that finite-amplitude MHD solutions exist for 
$\reym$ below its finite critical value. They associate this hysteresis with a 
nonlinear Lorentz force effect, which in this respect is quite similar to what 
occurs in the MRI dynamo problem. It is possible that the presence of strong
local shear in their model is responsible for subcriticality; however the
two types of subcritical dynamos look quite distinct at the moment.
To make another (far less rigorous) analogy with  hydrodynamic
transition problems,  their bifurcation diagram resembles that of the
subcritical bifurcation of Tollmien-Schlichting waves in plane Poiseuille 
flow (Zahn et~al. 1974 ; Herbert 1976 ; Orszag \& Patera 1980), which is a 
completely distinct phenomenon from the self-sustaining process in the 
same flow (Waleffe 2001).

\section{Subcritical shear dynamos and small-scale dynamo action\label{ssdynamo}}
A final point that is worth discussing is to what extent the self-sustaining 
coherent dynamo structures described in this paper are important to understanding
dynamo action in a highly turbulent medium. It has been shown that the 
hydrodynamic self-sustaining process is a cornerstone of transition to
turbulence in linearly stable shear flows and that this process leaves
an imprint on the statistical quantities (e.g. transport) associated
with the turbulent flow after the transition. As discussed by Lesur 
et~al. (2005), the hydrodynamic self-sustaining process in a shearing
box is a fundamentally lar\-ge-scale process that continuously
extracts energy from the shear (see their Fig.~9 and the corresponding
text). In other words, the SSP acts in the same way as a standard
linear instability  from the turbulence point of view, by forcing the
system at large scales. That nonlinear instabilities in shear flows 
extract energy from the shear in the same way as linear instabilities
extract energy from a general
free energy source is further supported by inspection of the energy budgets and of
the behaviour of structure functions (including those related to the forcing) in 
a turbulent flow driven by a linear instability like turbulent convection 
(Rincon 2006) and in a turbulent shear flow with no walls where the
transition process is fundamentally subcritical (Casciola
et~al. 2003).  The energy cascade clearly proceeds in a very similar
way for both types of forcing. 

We conjecture that self-sustaining MHD processes generated by subcritical shear
dynamos are also confined to large scales in the limit of large $\rey$,
and that their main role is to drive turbulence continuously by extracting
energy from the shear. If this were true, then an important consequence would
be that small-scale dynamo action should take place exactly in the same way in 
MRI turbulence with zero net flux and in turbulence driven by other means
(artificial forcing, thermal convection, MRI with 
net flux), provided that there is a sufficient scale separation between the 
forcing scales of the turbulence and the small-scale dynamo scales. 
There are now some clear numerical indications - including MRI dynamo 
simulations - that some universality with respect to the forcing
process exists for the kinematic stages of the small-scale dynamo at
$\prm > 1$. In this regime, the forementioned scale separation is easy
to obtain, because the  small-scale dynamo relies on the viscous scale
eddies (see Zel'dovich et~al. (1984) 
for theory and Schekochihin et~al. (2004) for an exhaustive numerical study). 
The numerical results obtained by Schekochihin  et~al. (2005) 
for idealized large-scale random forcing, by Christensen,  Olson \&  Glatzmaier (1999) 
and Cattaneo (2003) for convection and by Fromang et~al. (2007b) for a MRI dynamo set-up
all show a similar behaviour for the dynamo threshold in the $\rey-\reym$
plane. The results presented by Fausto Cattaneo at the recent Catania
workshop on MHD  also show that snapshots taken from turbulent
convection simulations and MRI dynamo simulations in a numerical
Taylor-Couette  experiment at large $\rey$ and $\reym$ are 
almost indistinguishable. The magnetic field 
maps at $\prm=1$ of Schekochihin et~al. (2004) and the ones by Fromang
et~al. (2007b) at $\prm=2$ (in the isotropic plane of their simulation
labelled $(x,z)$ in their notation, corresponding to $(-y,z)$ here)
also look very similar.
Overall, these new results tend to support our conjecture that there
is a large-scale subcritical process involving the MRI that drives
turbulence, and that this turbulence in turn operates as an 
independent small-scale dynamo at moderate to large $\prm$. We note in passing that 
the situation at low $\prm$ is more tricky  since it is currently unknown whether the 
small-scale dynamo in that regime has something to do with the forcing scales of the 
turbulence or if it is universal with respect to the forcing mechanism 
(Schekochihin et~al. 2007).

An important final remark regarding the MRI dynamo problem is that the
estimate for MRI growth rates $\gamma\sim \vec{V}_A\cdot\vec{k}$ predicts that 
even extremely small scales should be unstable to the MRI in the
presence of very weak fields, casting some doubt on the argument that
the MRI dynamo could be forced mostly at large scales
in the limit of large $\rey$ and $\reym$. We note that the
local MRI analysis, as any local analysis, is only valid when the
scale of the background field is  far larger than that of the
instability. In this respect, the previous growth rate estimate 
does not strictly apply at scales $1/k$ comparable to those of the strongly
tangled fields observed in MRI turbulence at moderate to large $\prm$,
thus there might well be some cut-off scale in the MRI dynamo problem
below which forcing by the MRI
becomes dynamically negligible. Fig.~4 of Fromang et~al. (2007b) 
shows that the forcing of poloidal magnetic fields in their simulations is 
fairly large-scale and falls off before the viscous scales. The numerics
are unfortunately not yet asymptotic and there is  no
published work on the MRI dynamo so far in which an appreciable scale 
separation between for\-cing and dissipation can be observed. It is 
therefore likely that testing our conjecture numerically and
discriminating between different scenarios will take a few more years.

\section{Conclusions}
In this paper, we discussed the concept of subcritical dynamo action
in shear flows and applied it to the problems of subcritical hydrodynamic 
transition and MRI dynamo action in accretion disks. We further showed
that the subcritical
dynamo scenario is relevant to many hydrodynamic and mag\-ne\-tohydrodynamic 
problems that involve two basic ingredients, namely shear 
and non-axisymmetric instabilities of shear-induced axisymmetric
toroidal velocity fieds or magnetic fields. We pointed out that the
coherence of the process
contrasts with the statistical description on which standard mean-field 
theory is based. We finally conjectured that coherent structures generated by 
subcritical dynamo action could be a backbone of MHD turbulence in shear flows
in the sense that their main role would be to extract energy from the shear to drive 
turbulence at large scales, thereby leaving some room in wavenumber space for 
an independent small-scale dynamo to proceed.

The whole picture is obviously not complete yet. There might be a way to unify
the statistical mean-field kinematic picture and the coherent subcritical picture.
There is a need to understand further which role SSPs play in
astrophysical objects such as accretion disks and stars. To this end,
there is a lot of work to do to relate the initial value  problem
description of these processes  to the phase space structure of the 
associated dynamical systems. We have shown that a description of 
subcritical MHD dynamos in terms of fixed points is helpful to 
understand simple configurations. However, numerical evidence (Cline
et~al. 2003) suggests that a fully chaotic behaviour can be obtained
easily for subcritical MHD dynamos in more complex 
configurations. Therefore, it might be necessary to describe these
dynamo processes in terms of more complicated phase space structures than
fixed points and to attempt to identify transition regions in parameter space
similar to the hydrodynamic ``edge of chaos'' (Schneider et~al. 2006, 
2007). From what we have learned so far, it is worth emphasizing that 
creating new connexions between the shear flow and transition 
community and the dynamo  community would undoubtedly prove extremely 
helpful to make some important progress on these matters.

\acknowledgements
We thank S\'ebastien Fromang, Geoffroy Le\-sur, 
John Papaloizou, Tobias Heinemann, Tarek You\-sef, Alexander 
Scheko\-chi\-hin, Tomo Tatsuno, Arnaud Antkowiak, Jean-Fran\c{c}ois
Pinton, Dan Lathrop, Hantao Ji, Fausto Cattaneo, Steve Tobias and
Sacha Brun for many interesting discussions on this topic. This work
has been partially  supported by the Leverhulme Trust and the Isaac
Newton Trust.

\end{document}